%% file: main.tex
\documentclass[sigconf,screen,authorversion,nonacm]{acmart} % sigconf
\AtBeginDocument{%
  }

\setcopyright{acmlicensed}
\copyrightyear{2025}
\acmYear{2025}

\setcopyright{rightsretained}
\acmConference[CHI PLAY Companion '25]{Companion Proceedings Annual Symposium on Computer-Human Interaction in Play}{October 13--16, 2025}{Pittsburgh, PA, USA}
\acmBooktitle{Companion Proceedings Annual Symposium on Computer-Human Interaction in Play (CHI PLAY Companion '25), October 13--16, 2025, Pittsburgh, PA, USA}
\acmDOI{10.1145/3744736.3749343}
\acmISBN{979-8-4007-2023-9/2025/10}

\usepackage{acronym}

\newacro{BAQ}{Brief Aggression Questionnaire}
\newacro{NASA TLX}{NASA Task Load Index}
\definecolor{ultrapurple}{HTML}{9B5DE5}

\usepackage{enumitem}

\newcommand{\mean}{\emph{M}=}
\newcommand{\sd}{\emph{SD}=}

\newcommand{\change}[1]{#1}

\renewcommand{\anon}[1]{<Anonymous for submission>}

\begin{document}

%%
%% The "title" command has an optional parameter,
%% allowing the author to define a "short title" to be used in page headers.
\title[How Casual Players Evaluate and Respond to Teammate Performance]{Beyond Competitive Gaming: How Casual Players Evaluate and Respond to Teammate Performance}
%\title[A Mixed-Methods Study Examining How Casual Gamers Evaluate and Respond to Teammate Performance]{"Laugh or Lash Out?": A Mixed-Methods Study Examining How Casual Gamers Evaluate and Respond to Teammate Performance}

%%
%% The "author" command and its associated commands are used to define
%% the authors and their affiliations.
%% Of note is the shared affiliation of the first two authors, and the
%% "authornote" and "authornotemark" commands
%% used to denote shared contribution to the research.
\author{Kaushall Senthil Nathan}
\authornote{\textbf{This is a pre-print version of our CHI PLAY 2025 paper. Please check out the DOI for the published version: \url{https://doi.org/10.1145/3744736.3749343}.}}
\authornote{Authors are also affiliated with the Department of Systems Design Engineering, University of Waterloo.}
\email{k3senthi@uwaterloo.ca}
\orcid{0009-0009-0846-1593}
\affiliation{
  \institution{HCI Games Group, Stratford School of Interaction Design and Business, University of Waterloo}
  \city{Waterloo}
  \state{ON}
  \country{Canada}
}

\author{Jieun Lee}
\authornotemark[1]
\email{jieun.lee@uwaterloo.ca}
\orcid{0009-0004-3121-4623}
\affiliation{
  \institution{Stratford School of Interaction Design and Business, University of Waterloo}
  \city{Waterloo}
  \state{ON}
  \country{Canada}
}

\author{Derrick M. Wang}
\authornotemark[1]
\email{dwmaru@uwaterloo.ca}
\orcid{0000-0003-3564-2532}
\affiliation{
  \institution{HCI Games Group, Stratford School of Interaction Design and Business, University of Waterloo}
  \city{Waterloo}
  \state{ON}
  \country{Canada}
}

\author{Geneva M. Smith}
\authornote{Also affiliated with the Department of Management Science and Engineering, University of Waterloo.}
\email{g38smith@uwaterloo.ca}
\orcid{0000-0002-6015-2589}
\affiliation{
  \institution{HCI Games Group, Stratford School of Interaction Design and Business, University of Waterloo}
  \city{Waterloo}
  \state{ON}
  \country{Canada}
}

\author{Eugene Kukshinov}
\email{eugene.kukshinov@uwaterloo.ca}
\orcid{0000-0002-3759-5218}
\affiliation{
  \institution{HCI Games Group, Stratford School of Interaction Design and Business, University of Waterloo}
  \city{Waterloo}
  \state{ON}
  \country{Canada}
}

\author{Daniel Harley}
\email{dharley@uwaterloo.ca}
\orcid{0000-0003-4309-584X}
\affiliation{
  \institution{Stratford School of Interaction Design and Business, University of Waterloo}
  \city{Waterloo}
  \state{ON}
  \country{Canada}
}

\author{Lennart E. Nacke}
\email{lennart.nacke@acm.org}
\orcid{0000-0003-4290-8829}
\affiliation{%
    \institution{HCI Games Group, Stratford School of Interaction Design and Business, University of Waterloo}
    \city{Waterloo}
    \state{ON}
    \country{Canada}
}

%%
%% By default, the full list of authors will be used in the page
%% headers. Often, this list is too long, and will overlap
%% other information printed in the page headers. This command allows
%% the author to define a more concise list
%% of authors' names for this purpose.
\renewcommand{\shortauthors}{Senthil Nathan et al.}

%%
%% The abstract is a short summary of the work to be presented in the
%% article. 
\begin{abstract}
Teammate performance evaluation fundamentally shapes intervention design in video games. However, our current understanding stems primarily from competitive E-Sports contexts where individual performance directly impacts outcomes. This research addresses whether performance evaluation mechanisms and behavioural responses identified in competitive games generalize to casual cooperative games. We investigated how casual players evaluate teammate competence and respond behaviourally in a controlled between-subjects experiment (N=23). We manipulated confederate performance in Overcooked 2, combining observations, NASA TLX self-reports, and interviews. We present two key findings. (1) Observations revealed frustration behaviours completely absent in self-report data. Thus, these instruments assess fundamentally distinct constructs. (2) Participants consistently evaluated teammate performance through relative comparison rather than absolute metrics. This contradicts task-performance operationalizations dominant in competitive gaming research. Hence, performance evaluation frameworks from competitive contexts cannot be directly applied to casual cooperative games. We provide empirical evidence that performance evaluation in casual games requires a comparative operationalization.
\end{abstract}

%%
%% The code below is generated by the tool at http://dl.acm.org/ccs.cfm.
%% Please copy and paste the code instead of the example below.
%%


%%
%% Keywords. The author(s) should pick words that accurately describe
%% the work being presented. Separate the keywords with commas.

%\received{20 February 2007}
%\received[revised]{12 March 2009}
%\received[accepted]{5 June 2009}

%%
%% This command processes the author and affiliation and title
%% information and builds the first part of the formatted document.
\maketitle
\twocolumn
\input{Chapters/1-Introduction}
\input{Chapters/2-Related_Works}
\input{Chapters/3-Methods}
\input{Chapters/4-0-Results}

\input{Chapters/5-Discussion}

\begin{acks}
This study was supported by the SSHRC INSIGHT Grant (grant number: 435-2022-0476), Dr. Nacke's NSERC Discovery Grant (grant number: RGPIN-2023-03705), Dr. MacArthur's NSERC Discovery Grant (grant number: RGPIN-2024-06734), Canadian Foundation for Innovation John R. Evans Leaders Fund (CFI JELF) (grant number: 41844), Mitacs Accelerate Grant (grant number: IT40801), Meta Research Award, Lupina Foundation Postdoctoral Research Fellowship, and the Provost’s Program for Interdisciplinary Postdoctoral Scholars at the University of Waterloo. We would also like to thank our lab members and the participants of this study for their time and assistance with this project.
\end{acks}

\balance
\bibliographystyle{ACM-Reference-Format}
\bibliography{references}

\end{document}

%% file: Chapters/1-Introduction.tex
\section{INTRODUCTION}
\label{sec:intro}

Teammate performance fundamentally shapes player experiences and behavioural outcomes in multiplayer video games. It influences everything from team success to player retention~\cite{musick_leveling_2021, gisbert-perez_key_2024}. Research in this domain has predominantly focused on competitive, high-stakes E-Sports contexts~\cite{martoncik_e-sports_2015, sachan_social_2025, banyai_psychology_2019}, where performance directly correlates with professional outcomes. These studies have established that increased teammate performance is related to team success and player enjoyment~\cite{bonilla_gorrindo_psychological_2022, hanghoj_esports_2019, musick_leveling_2021}, while decreased performance can contribute to toxicity and frustration~\cite{shen_viral_2020, mclean_toxic_2020}. Consequently, substantial effort has been devoted to operationalizing teammate performance in competitive contexts~\cite{sabtan_current_2022, gisbert-perez_key_2024}, particularly in games like \textit{League of Legends} and \textit{Valorant}~\cite{abbott_perceptions_2022, lee_characterizing_2024, maharani_understanding_2024}.

However, this focus on competitive games has left a considerable knowledge gap regarding teammate performance in casual cooperative games where player motivations, stakes, and interaction patterns differ substantially~\cite{trotter_social_2021, martoncik_e-sports_2015}. \change{For the purposes of this study, we define casual cooperative games as non-competitive games where the roles and duties of the players are dependent on each other to complete the objectives of the games. Popular examples include \textit{Overcooked} and \textit{It Takes Two}, where players must work together to complete in-game tasks like meal preparation and puzzle solving respectively~\cite{noauthor_overcooked_nodate, HazelightStudios_2021}.} While performance effects remain relevant in casual contexts~\cite{Wang_2025, greitemeyer_theres_2013}, the operationalization strategies and behavioural frameworks developed for competitive games may not translate directly to cooperative casual gaming environments. Without understanding how casual players evaluate competence and respond to teammate performance, game designers cannot create appropriate feedback systems, and researchers cannot develop effective interventions to promote positive cooperative behaviours. This could lead to misapplied research frameworks that fail to improve player experiences in the growing casual multiplayer gaming market.

This Work-in-Progress addresses this problem by investigating teammate performance evaluation and behavioural responses in \textit{Overcooked 2}~\cite{noauthor_overcooked_nodate}. We conducted a controlled, between-subjects experiment with 23 participants across competent and incompetent teammate conditions. The same confederate played as the participants’ teammate in both conditions. We used mixed methods including physiological measures, questionnaires, behavioural observation, and semi-structured interviews to examine the following research questions:

\begin{itemize}
    \item[\textbf{RQ1:}] How do players evaluate teammate performance in casual games?
    \item[\textbf{RQ2:}] What impact does teammate performance have on a player’s frustration?
    \item[\textbf{RQ3:}] What impact does teammate performance have on a player’s cooperative behaviour?
\end{itemize}

Given the unknowns about how measures from E-Sports might apply to this context, we used a number of strategies for data collection, including an Electroencephalogram (EEG) to assess mental workload during gameplay, a post-game self-assessment with the \ac{NASA TLX} to assess frustration, followed by a short semi-structured interview to better understand participants’ choices and experiences during play. We also video- and audio-recorded participants’ play sessions to observe and assess their cooperative behaviour. 

We found promising areas for future research in cooperative casual games. Specifically, we need to examine how frustration is measured and how players define competence \change{within this context}. For example, our observational measures captured frustration levels that were not identified by our self-report tool---most likely due to differing definitions \change{underlying the tools}---and indicating that the measures assess distinct constructs. The \ac{NASA TLX} focuses on emotional states (e.g., irritation, stress), while our observational assessment of \textit{in-game frustration}, as defined by \citet{gilleade_using_2004}, focused on confusion about overcoming a challenge. Future work measuring frustration needs to be more precise. Researchers should define frustration consistently across all tools, or use diverse assessments for a more detailed understanding. Similarly, while operationalization of teammate competence used in competitive games (e.g., task performance) is crucial in professional contexts, our participants’ perceptions of competence instead foregrounded experiential and cooperative factors. Participants often described their performance in relative terms (i.e., in relation to that of their teammate), rather than in terms of their task accuracy or success rate. These results challenge existing operationalization frameworks and provide foundational insights to develop and customize performance evaluation methods \change{for casual games}.

%% file: Chapters/2-Related_Works.tex
\section{RELATED WORK}
\label{sec:relatedwork}

\subsection{Operationalization of Teammate Performance}

There is no consensus on how to measure teammate performance in video games. Unsurprisingly, the E-Sports research community has been leading the effort to better understand teammate performance, striving to better understand how to train players~\cite{sabtan_current_2022}. However, the operationalizations vary greatly. \citet{gisbert-perez_key_2024}, in their systematic review, found that performance can be viewed as a result of outcomes (win, loss, or tie/draw) or actions (i.e., task performance or behaviour). The first form of assessment is best illustrated in \citet{trepanowski_sexism_2024}, where participants rated the performance of a streamer on a 1 to 7 scale after viewing a video clip of gameplay. With some similarities between E-Sports and traditional sports, this perspective is analogous to sports networks discussing players' statistics after the match~\cite{pizzo_esports_2022}. While this approach can be easily applied across games, it does not capture what constitutes good or bad performance \change{during the game}. Therefore, before testing its effects, we need to understand what constitutes good performance. 

The second form of assessment put forward by \citet{gisbert-perez_key_2024}, where performance is viewed as a result of actions, can be further divided into task and contextual performances. Task performance refers to a numerical measure of task completion or accuracy, best illustrated in \citet{neri_personalized_2021}, who used the ratio of enemies killed to player death to measure how well a player was trained. Contextual performance refers to behaviours of a player towards success in the game, which ranges from their communication with their teammates to their diets and training regimens~\cite{baumann_exploratory_2024}. While it may seem intuitive to evaluate performance using both task and contextual performance, different games require vastly different competencies~\cite{nagorsky_structure_2020}. Therefore, contextual performance may vary greatly across games and genres. This is contrasted by task performance being generalized more easily across games (e.g., the Kill-Death ratio used by \citet{neri_personalized_2021} can be used in other First-Person Shooters), and possibly even genres. However, before we commit to this assessment, we must be mindful that all of the previous studies only examined competitive games, which may not directly apply to casual games because motivations for competitive and casual players differ greatly~\cite{martoncik_e-sports_2015}. \change{Therefore, we need to evaluate whether these prior assessments of task performance are valid in casual games.}

\subsection{The Effects of Teammate Performance are Understudied in Casual Games}

Existing research on teammate performance has focused on popular competitive titles with massive player bases, such as \textit{League of Legends}~\cite{mclean_toxic_2020}, \textit{Valorant}~\cite{maharani_understanding_2024}, and \textit{Counter Strike: Global Offensive}~\cite{trepanowski_sexism_2024}. Not only is comparative research on casual cooperative games scarce, cooperative games are more often evaluated as tools or interventions~\cite{hanghoj_can_2018, greitemeyer_theres_2013}, rather than environments. For example, \citet{hanghoj_can_2018} used a casual cooperative game to test its effectiveness as a tool in teaching math. However, current research aiming to reduce negative behaviour in games by identifying their causes is also relevant for assessing teammate environments and behaviours in casual games. For example, \citet{shen_viral_2020} conducted a longitudinal study of toxicity in \textit{World of Tanks}, a competitive team shooter, and found that high skill disparities, among other factors, lead to more instances of frustration and toxicity. In a similar vein, \citet{mclean_toxic_2020} found that in \textit{League of Legends}, frustration was most pronounced in the absence of a teammate's expected cooperative behaviour. This research has informed interventions to prevent negative behaviour in games~\cite{bongaards_personalized_2024, reid_feeling_2022}. However, the focus on competitive games overshadows the casual cooperative genre (e.g., \textit{Overcooked 2}~\cite{noauthor_overcooked_nodate}) which makes up a sizable portion of the player population. There have been studies that focus on the cooperative multiplayer genre, but still rely on the competitive aspects the game offers. For example, ~\citet{breuer_sore_2015} found that, in a co-located, competitive game played in a casual context, trash-talking had no effect on participants' frustration, which may suggests there's more complex dynamics at play in non-professional contexts. The broader lack of attention to casual cooperative games suggests the need for research to examine the possible effects of team performance to create interventions that promote positive behaviours similar to those seen in competitive games. 

%% file: Chapters/3-Methods.tex
\section{METHODS}
\label{sec:meth}

We conducted a between-subjects, randomized controlled experiment with two conditions: Competent (C) and Incompetent (IC). \change{Author 4 and a volunteer from our research lab} (``the confederate'') acted as the teammate for each participant. \change{While the anonymous nature of the confederate may not capture the range of relationships seen in cooperative players, we chose this approach because it provided consistent benchmarks, communication instructions, and behaviour across the conditions for the confederates.} In the Competent condition, the confederate would not make any mistakes. In the Incompetent condition, the confederate intentionally fails 25\% of tasks in the first two levels, 50\% in the next two, and 67\% of tasks in the last two levels. To test participant reactions to varying failure levels, we kept the confederate's error rate believable during the initial, easier levels. The confederate was also instructed to keep responses brief and only respond when asked a question to not lead the participants. Ethics approval was obtained from a Research Ethics Board at the University of Waterloo (\#46330).

We selected \textit{Overcooked 2} because the game's level design requires varying degrees of cooperation, and its two-player design reduced the complexity of team dynamics and confounding variables larger team games may have.

\subsection{Measures}\label{sec:measures}

Before gameplay, we collected each participant's age, gender, race, video game types played, hours per week spent playing video games, and whether they preferred challenges while playing video games. They completed all questionnaires electronically via a computer in the experiment room. We used the \ac{BAQ}~\cite{webster_brief_2014} to rule out predispositions to frustration or anger across conditions.

%Interview%
\subsubsection{Semi-Structured Interview}
Following gameplay, participants completed one-on-one, 15-minute, semi-structured interviews. We asked open-ended questions to gauge their views on their teammate's performance, their in-session behaviour, and whether their performance conceptualization aligned with our operationalization post-debriefing (\textbf{RQ1}). At the end of the interview, we told the participants about the confederate and that they were instructed to play a certain way, and we asked whether this knowledge changed their perspective of their experience.

\subsubsection{Frustration and Cooperation}
To investigate RQ2, we used the \ac{NASA TLX}~\cite{Hart_Field} to measure the participant's subjective frustration. The \ac{NASA TLX} consists of six dimensions: \textit{Mental Workload}, \textit{Physical Demand}, \textit{Temporal Demand}, \textit{Performance}, \textit{Effort}, and \textit{Frustration}. \change{We were mostly interested in \textit{Frustration} because it is most relevant to our research question and because research connecting the remaining constructs to either frustration or cooperation in video games is scant. Despite this, we administered the full questionnaire, following best practices to maintain experimental validity~\cite{bustamante_measurement_2008, Kukshinov_Tu_Szita_SenthilNathan_Nacke_2025, Perrig_Aeschbach_Scharowski_vonFelten_Opwis_Brühlmann_2024}}.

Surveys and interviews are frequently used to measure outcomes of poor teammate performance, especially toxicity or frustration~\cite{frommel_toxicity_2024, wijkstra_fighting_2024, zhang_toxicity_2024, nexo_toxic_2024, trepanowski_sexism_2024}. To avoid the limitations of only relying on self-reported measures, we also observed players' gameplay and used a physiological measure of frustration. For observation, the participant's voice and facial reactions were recorded using a microphone and a webcam, respectively, alongside the gameplay to capture the frequency of frustration and cooperative behaviour. In addition to this, we chose a single channel, in-ear Electroencephalogram (EEG) device manufactured by \textit{IDUN Technologies}, given that EEGs have previously been used in video game research~\cite{ahonen_electroencephalography_2021}, to be our physiological measure for frustration. These measures allow us to capture frustration and cooperative behaviour over time.
\subsection{Sample Characteristics}
We recruited participants by advertising the study using departmental newsletters and posters distributed to classes held on campus. Any participant between the ages of 18 to 65 was eligible to participate, provided they do not meet any of the following exclusion criteria: (1) have a history of violence or marked persistent excessive anxiety in response to acute mental stress, (2) use an implanted device (such as a pacemaker), (3) have an allergy to silver, and (4) if they have any condition that prevents them from playing a video game. \change{Items (2) and (3) were added according to the EEG's usage guidelines.}

Twenty-three participants were recruited for this study (13 self-identified as female, 10 as male). Participants' ages ranged from 20 to 23 years (\mean{21.39}, \sd{0.94} years). Fourteen participants self-identified as East Asian, three as Southeast Asian, three as South Asian, three as White, and one as Biracial (East Asian and Southeast Asian). The hours played per week ranged from 0 to 10 (\mean{4.89}, \sd{3.63} hours). Participants reported a wide range of games and genres played, ranging from competitive videos games like \textit{Call of Duty}, to casual games like \textit{Stardew Valley}. All participants reported that they did not know the confederate personally. A Wilcoxon signed-rank test indicated that trait aggression is not significantly different across conditions ($W = 52.5$, $p = 0.422$), confirming that it is not a confounding factor in this experiment.

\subsection{Procedure}
\label{sec:procedure}

In the study room, \change{the first author} presented the participant with the deceptive consent form, reiterated the exclusion criteria, and obtained consent. Then, they asked the participant to fill out the demographics questionnaire, and recorded a baseline reading of EEG data to capture low and high workload (similar to \citet{cirett_galan_eeg_2012}). They also took note of the signal quality of the EEG and made necessary adjustments.

To standardize the study procedure and participant motivation, the researcher told the participant not to repeat any levels and to try and get three stars in all six levels (i.e., a high score). Furthermore, to ensure the participant would speak to the confederate, we instructed them to act as a leader, and discuss strategy with the other player. Following this, the researcher reconfirmed with the participant before starting the video recording and voice call between the participant and the confederate. Finally, the researcher retreated to the observation space with a one-way mirror. 

Participants then played through a tutorial and the first six levels of the game. The tutorial was included to ensure that participants new to the game could play. Once they finished playing through all the levels, the researcher returned to the room, terminated the voice call and the recording, and administered the post-experiment questionnaires containing the \ac{BAQ} and the \ac{NASA TLX}. Finally, the researcher conducted the one-on-one semi-structured interview. Participants were remunerated with \$15 CAD.

\subsection{Analysis}
\label{sec:analysis}

\subsubsection{Quantitative Analysis}
\label{sec:quan_anal}

For our questionnaire data, we began with the Shapiro-Wilk test. As normality assumptions were not met, we used the Mann-Whitney U test as an alternative. For the video recordings, we performed a content analysis to identify key behaviours related to our RQs. \change{Specifically, we analyzed moments when participants verbally communicated with the confederate.} \change{The second and third authors} independently analyzed the videos based off a code book consisting of items from two relevant sources (\textit{At-Game Frustration} caused by controls and \textit{In-Game Frustration} caused by game mechanics~\cite{gilleade_using_2004}, \textit{Laughing together}, \textit{Worked out strategies}, \textit{Helping each other}, \textit{Global strategies}, \textit{Waited for each other}, \textit{Got in each others' way}, and \textit{Other cooperative behaviour}~\cite{seif_el-nasr_understanding_2010}), while allowing new entries relevant to our RQs to be added. \change{The full code book is attached in the supplementary material.} We calculated inter-rater reliability and resolved any conflicts with \change{the first author} as a tie-breaker ~\cite{drisko2016content}. The researchers summed the codes and calculated significance in the same manner as our questionnaire data. We cleaned the EEG data and removed any artifacts, then applied a band-pass filter of 1-50 Hz. Next, we used a Fast Fourier Transformation to transform the EEG data from time-domain to frequency-domain. Lastly, we compared the frequency observed in different bands across groups to determine if there is a statistically significant difference.

\subsubsection{Qualitative Analysis} 
\label{sec:qual_anal}

We transcribed the interview audio using \textit{Otter.ai} \footnote{\url{https://otter.ai}} and corrected any inaccurate transcriptions and removed filler words. To analyze this data, we employed a reflexive thematic analysis, following the six phase analytical process outlined by \citet{Braun_Clarke_2019}\change{, because it aligned with our interpretive, inductive approach for the interview data}. First, all three researchers involved in the analysis individually familiarized with the data. In \change{an} iterative process, \change{the first and third authors inductively} coded five interview transcripts (\~20\%) at a time to extract information relevant to our research questions, generating descriptive codes (e.g., `camaraderie reduces frustration'). \change{After each round, the coders met to compare interpretations. This was not done to reach consensus, but rather to deepen reflexive insight through our discussion~\cite{Braun_Clarke_2021}. \change{The second author}, not involved in the initial coding, later reviewed the full dataset to provide a fresh perspective and support reflexive quality checking of the final themes.} Following the third iteration, we developed initial themes through affinity mapping. After coding all the data, all three researchers reviewed the codes to ensure that they stayed true to the meaning of the data. Finally, we reviewed and refined the themes.

%% file: Chapters/4-0-Results.tex
\section{Preliminary Results}
\label{sec:results}

\input{Chapters/4-1-QuanResults}

\input{Chapters/4-2-QualResults}

%% file: Chapters/4-1-QuanResults.tex
\subsection{Quantitative Findings}

Given the small sample size, the Shapiro-Wilk test was performed on the frustration data in order to test the assumption of normality (\textit{W} = 0.85). As the assumption of normality was violated, a Wilcoxon signed-rank test was used (\(M_C=\) 25.42, \(SD_C=\) 23.4; \(M_{IC}=\) 30, \(SD_{IC}=\) 23.87, \textit{W} = 56). 

We discarded the EEG data as the recordings proved to be highly unstable, with the signal quality, taken at the beginning, of 15 recordings (out of 23) below the manufacturer's standard. In these cases, efforts were made to increase signal quality according to manufacturer's instructions (e.g., re-cleaning the electrodes), but signal quality remained poor. More robust testing is needed to understand \change{the causes of this failure and} whether similar EEGs can be used in this setting.

\begin{figure*}
    \centering
    \includegraphics[width=\textwidth]{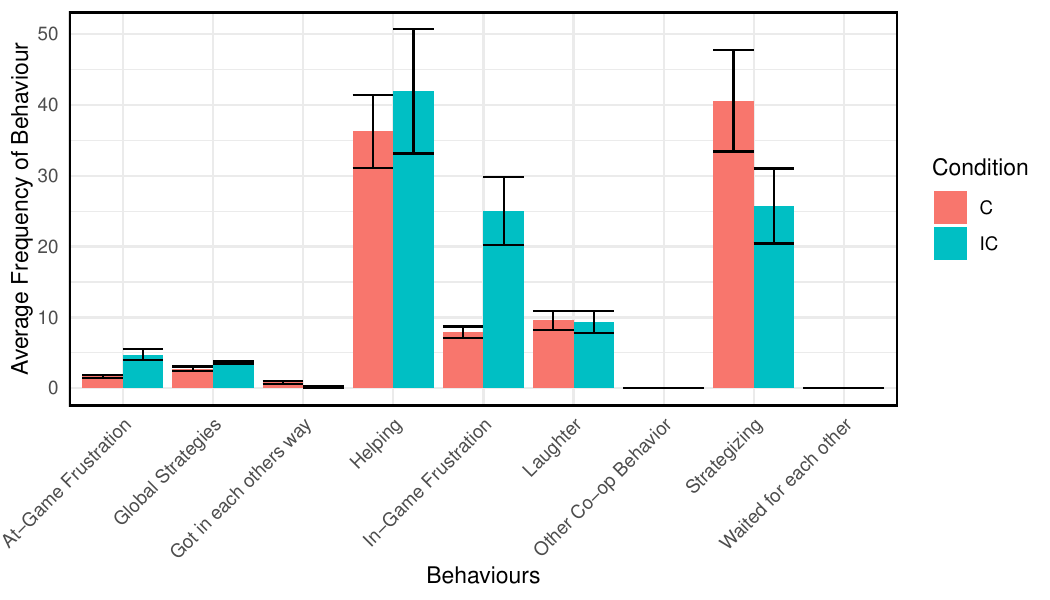}
    \caption{Average Frequency of Behaviour across Competent and Incompetent Conditions}
    \Description{This figure visualizes the average frequency of behaviour across Competent and Incompetent conditions.}
    \label{fig:frequency}
\end{figure*}

For the quantitative content analysis, while inter-rater reliability was high (Initial Cohen's Kappa Coefficient = 0.48), we are aware that with our small sample size, these results may be underpowered. Therefore, we highlight the notable trends we observed while acknowledging the limitations of the current dataset. While the average occurrence of \textit{At-Game Frustration} behaviours per participant was low for both conditions (\(M_C=\) 1.667, \(SD_C=\) 0.188; \(M_{IC}=\) 4.72, \(SD_{IC}=\) 0.76), \textit{In-Game Frustration} behaviours differed vastly across conditions (\(M_C=\) 7.9, \(SD_C=\) 0.8; \(M_{IC}=\) 25, \(SD_{IC}=\) 4.79). Regarding the cooperative behaviours examined, only the average occurrence of \textit{Worked out Strategies} had a high value and a substantial difference across conditions as well (\(M_C=\) 40.58, \(SD_C=\) 7.16; \(M_{IC}=\) 25.73, \(SD_{IC}=\) 5.28). These results are outlined in \autoref{fig:frequency}. These results show two things. First, when frustration is broken down into \textit{At-Game} and \textit{In-Game}, a difference in frequency of \textit{In-Game Frustration} indicates a difference and a potential cause for the frustration that the NASA TLX did not capture. Second, not all aspects of cooperative behaviour in our code book ~\cite{seif_el-nasr_understanding_2010} are affected equally by our manipulation of confederate competence, indicating unknown causes influencing the participant's various cooperative behaviour (e.g., different frequency to strategize across conditions, versus roughly equal frequency of laughter across conditions).

%% file: Chapters/4-2-QualResults.tex
\subsection{Qualitative Findings}

We developed two main themes that describe the participants' interpretations of teammate performance and its effects on frustration and/or cooperative behaviour. 

\subsubsection*{\textbf{Theme 1: Participants evaluate a teammate's performance relative to their own}}

While E-Sports studies often strive for quantitative measures like task accuracy to evaluate teammate performance~\cite{mclean_toxic_2020, neri_personalized_2021}, our participants' responses suggest that perceived differences between performance mattered most. When asked about the confederate's skill, the participants always compared it with their own (e.g., ``I was probably playing fairly poorly as well.'' (\textbf{P7})). This trend of comparing personal performance with the confederate extended to aspects beyond task performance, including team behaviour or attitude. As \textbf{P19} stated, ``I was very calm, and [the confederate was] very calm. We just talked to each other.'' This illustrates that, regardless of whether it was a task performance measure or a contextual attitude, the participant consistently assessed performance in a relational way. It is important to acknowledge that these conclusions may only apply to casual multiplayer games, as participants considered competitive games different, stating that ``one bad player in a team can kind of ruin everybody.'' (\textbf{P20}), supporting the claim that casual, cooperative games provide an equal playing field where players contribute to a collective goal~\cite{baek_toward_2022}. Thus, we surmise that, in casual cooperative games, teammate performance must be evaluated in relative terms in addition to other quantitative measures to increase external validity.

\subsubsection*{\textbf{Theme 2: Frustration and cooperative behaviours are most influenced by the teammate, not the game}}

Consistent with existing literature~\cite{Wang_2025, shen_viral_2020, mclean_toxic_2020}, participants believed that \change{their overall feeling of frustration during the game stemmed not only from their own performance, but also that of the teammate (the confederate)}. For example, one participant playing with an incompetent teammate reported that failing to achieve a task performance goal caused their frustration: ``the last round we didn't get the three stars. So that was, of course, upsetting'' (\textbf{P3}). With regards to cooperative behaviour, another participant reported that the responsiveness of the confederate (i.e., high contextual performance) encouraged them to strategize better and mitigate some frustration that arose from playing in the incompetent condition: ``we experienced some frustrations as the levels progressed but I think that was smoothed out as we figured out how to work past that.'' (\textbf{P18}). Regarding the effect of the casual game on participants' behaviour, some participants considered the game design as potentially adding to their frustration: ``nearing the end as the timer was going down ... it's like a bit more pressure.'' (\textbf{P13}), while others claimed that the game design reduced frustration (``Overcooked is a fun party game, so I'm not stressing out.'' (\textbf{P25})). Again, participants saw this as different from competitive games, where players ``play against an opponent [and] get aggressive [towards their teammates].'' (\textbf{P1}). Here, even repeated failures did not immediately cause frustration: (``if they keep doing stuff wrong, I may say something'' (\textbf{P24})). These results suggest that teammate performance does have an effect on their frustration and cooperative behaviour, while the game itself has no consistent effect across the sample.

%% file: Chapters/5-Discussion.tex
\section{DISCUSSION}
\label{sec:discuss}

In this study, we examined whether the operationalization of teammate performance from previous research (including the finding that poor teammate performance causes frustration) can be applied in the context of a casual, cooperative game. We tested this in a controlled experiment using qualitative and quantitative measures. With our preliminary results, we present two main findings. Firstly, we identified shortcomings in the operationalization of teammate competence for casual players that can limit external validity. Secondly, observational measures of frustration may capture behaviours absent in self-reports, likely because the two assess distinct frustration constructs. 

\subsubsection*{\textbf{RQ1: How do players evaluate teammate performance in casual games?}}

Our interview results suggest that the quantitative operationalization of teammate performance seen in previous research~\cite{mclean_toxic_2020} falls short in casual games, and so does the holistic approach seen in E-Sports~\cite{Bonilla_Gorrindo_Chamarro_Ventura_2022}. Instead, all our participants compared their performance to that of the confederate to determine how well they did. This extended to both task performance and contextual measures of performances. While uncommon, some studies, like \citet{shen_viral_2020}, have used skill disparities as a behavioural predictor in games. This also aligns with existing work, including gamification work stating that comparative leaderboards have more influence than absolute leaderboards~\cite{Bai_Hew_Gonda_2021}, and exergaming literature showing players gaining elevated motivation through comparing their performance with their teammates~\cite{Wang_2025}. As further research into casual games unfolds, \textbf{we recommend the use of comparative and relational measures of performance (e.g., the confederate played better, worse, or the same as the participant in this aspect) to increase external validity.}

\subsubsection*{\textbf{RQ2/3: What is the effect of a teammate's performance on the player's frustration and cooperative behaviour?}}

A key finding in our interviews was that the casual game mechanics contributed to a lack of frustration as well as an environment where frustrations could be dismissed or easily overcome through communication. However, our observations also captured frustration that was not identified by our self-report tool. We believe this is not a result of social-desirability bias or memory effect, as our interview results align with our observations. Therefore, we are inclined to attribute this to the differing definitions in the measures (e.g., the \ac{NASA TLX} defines the frustration question as ``How insecure, discouraged, irritated, stressed and annoyed were you?''~\cite{Hart_Field}, while the observational measure proposed by \citet{gilleade_using_2004} defined \textit{In-Game Frustration} as ``that which arises from a failure to know how a challenge is to be completed.'' Our observation of cooperative behaviours showed that only one behaviour (i.e., \textit{Worked out strategies}) substantially differed across conditions, with some aspects (i.e. \textit{Helping each other}, \textit{Global strategies}, \textit{Waited for each other}) not directly observed but mentioned by participants in the interview. One possible reason for the difference in \textit{Worked out strategies} is that the player could rely more on the `competent' teammate and thus were more likely to strategize with them, indicating nuances in the cooperative behaviour based on the traits of the team. Therefore, we suggest that future research \textbf{define frustration and cooperative behaviour so that both demonstrate strong construct validity and can be applied across different types of measures.}

\subsection{Limitations and Future Work}

As a Work-in-Progress, our study had several limitations. Firstly, the small sample of this study made it difficult to conclude anything definitive about \textbf{RQ2} and \textbf{RQ3}. While expected, our in-lab experiment setup which included observation, recording, and a stranger as a co-player made a few participants feel uneasy (e.g., they tried to act nice to the confederate despite feeling frustrated), which could reduce the validity of our observational measure. Finally, the failure of the EEG in our research raises questions regarding the feasibility of the device in this use case. 

There are several interesting directions that future research can explore. First, expanding the sample size with the current methodology to determine if the difference found by the questionnaire and observational measures have a statistically significant difference, with a pivot to include a comparative measure rather than an absolute one. Furthermore, with teammate performance viewed in comparative terms in \textit{Overcooked 2}, team dynamics in other casual games can be examined and interventions can use comparative performance to identify where performance diverges and how to correct it. \change{Expanding on this, researchers can also look at different player traits, such as the \textit{Online Gaming Motivation} scale~\cite{10.1145/2207676.2208681} or the participants preference for challenge (which we  collected but did not use in the results due to a low sample size), and types of relationships between casual players (e.g., does this trend differ when played with friends/family).} 

We also highlight the nuances of cooperation and frustration that were not explored in this study. For example, future work could extend our experimentation by exploring the potential conflation between the frequency of cooperative behaviour and frustration. Another possible direction to build off this WiP is the need for either a synthesis or selection of definitions for frustration and cooperative behaviour for games research. Future research needs to identify what frustration and cooperative behaviour mean in casual games and whether that applies across different types of measures. For example, in this study, we measured vocal frustration through observation and self-report measures, but these are far from the only ways frustration can be measured. Two possible ways this can be achieved is by (1) examining a large sample of casual gamers to compare different definitions of frustration and cooperative behaviour, or (2) synthesizing the existing literature to identify definitions that can be applied across measurement types. 

\subsection{Conclusion}

This Work-in-Progress responds to the current emphasis on competitive gaming in the literature by investigating how teammate performance and its effects are understood in casual cooperative games. Through a controlled experiment and interviews, we examined whether existing operationalizations of performance and frustration, developed in competitive games, apply to casual players. Our preliminary results present two key findings. First, participants evaluated teammate performance in relative terms, comparing the confederate’s performance to their own, rather than relying on a fixed, absolute standard. Second, our observations of frustration revealed behaviours not captured by self-reports, likely because the two measures assess fundamentally different constructs. With this, we put forward one recommendation per finding respectively: (1) as further research into casual games unfolds, we recommend the use of comparative measures of performance to increase external validity, and (2) we suggest future games research put forward a definition of frustration and cooperative behaviour that demonstrate strong construct validity and that can be applied across different types of measures.

%% file: main.bbl
%%% -*-BibTeX-*-
%%% Do NOT edit. File created by BibTeX with style
%%% ACM-Reference-Format-Journals [18-Jan-2012].

\begin{thebibliography}{47}

%%% ====================================================================
%%% NOTE TO THE USER: you can override these defaults by providing
%%% customized versions of any of these macros before the \bibliography
%%% command.  Each of them MUST provide its own final punctuation,
%%% except for \shownote{}, \showDOI{}, and \showURL{}.  The latter two
%%% do not use final punctuation, in order to avoid confusing it with
%%% the Web address.
%%%
%%% To suppress output of a particular field, define its macro to expand
%%% to an empty string, or better, \unskip, like this:
%%%
%%% \newcommand{\showDOI}[1]{\unskip}   % LaTeX syntax
%%%
%%% \def \showDOI #1{\unskip}           % plain TeX syntax
%%%
%%% ====================================================================

\ifx \showCODEN    \undefined \def \showCODEN     #1{\unskip}     \fi
\ifx \showDOI      \undefined \def \showDOI       #1{#1}\fi
\ifx \showISBNx    \undefined \def \showISBNx     #1{\unskip}     \fi
\ifx \showISBNxiii \undefined \def \showISBNxiii  #1{\unskip}     \fi
\ifx \showISSN     \undefined \def \showISSN      #1{\unskip}     \fi
\ifx \showLCCN     \undefined \def \showLCCN      #1{\unskip}     \fi
\ifx \shownote     \undefined \def \shownote      #1{#1}          \fi
\ifx \showarticletitle \undefined \def \showarticletitle #1{#1}   \fi
\ifx \showURL      \undefined \def \showURL       {\relax}        \fi
% The following commands are used for tagged output and should be
% invisible to TeX
\providecommand\bibfield[2]{#2}
\providecommand\bibinfo[2]{#2}
\providecommand\natexlab[1]{#1}
\providecommand\showeprint[2][]{arXiv:#2}

\bibitem[Abbott et~al\mbox{.}(2022)]%
        {abbott_perceptions_2022}
\bibfield{author}{\bibinfo{person}{Callum Abbott}, \bibinfo{person}{Matthew Watson}, {and} \bibinfo{person}{Phil Birch}.} \bibinfo{year}{2022}\natexlab{}.
\newblock \showarticletitle{Perceptions of {Effective} {Training} {Practices} in {League} of {Legends}: {A} {Qualitative} {Exploration}}.
\newblock \bibinfo{journal}{\emph{Journal of Electronic Gaming and Esports}} \bibinfo{volume}{1}, \bibinfo{number}{1} (\bibinfo{date}{Oct.} \bibinfo{year}{2022}), \bibinfo{numpages}{11}~pages.
\newblock
\urldef\tempurl%
\url{https://doi.org/10.1123/jege.2022-0011}
\showDOI{\tempurl}


\bibitem[Ahonen et~al\mbox{.}(2021)]%
        {ahonen_electroencephalography_2021}
\bibfield{author}{\bibinfo{person}{Ville Ahonen}, \bibinfo{person}{Marko Leino}, {and} \bibinfo{person}{Tarmo Lipping}.} \bibinfo{year}{2021}\natexlab{}.
\newblock \showarticletitle{Electroencephalography in {Evaluating} {Mental} {Workload} of {Gaming}}. In \bibinfo{booktitle}{\emph{2021 43rd {Annual} {International} {Conference} of the {IEEE} {Engineering} in {Medicine} \& {Biology} {Society} ({EMBC})}}. \bibinfo{publisher}{IEEE}, \bibinfo{address}{New Jersey, USA}, \bibinfo{pages}{845--848}.
\newblock
\urldef\tempurl%
\url{https://doi.org/10.1109/EMBC46164.2021.9629772}
\showDOI{\tempurl}
\newblock
\shownote{ISSN: 2694-0604}.


\bibitem[Baek et~al\mbox{.}(2022)]%
        {baek_toward_2022}
\bibfield{author}{\bibinfo{person}{In-Chang Baek}, \bibinfo{person}{Tae-Gwan Ha}, \bibinfo{person}{Tae-Hwa Park}, {and} \bibinfo{person}{Kyung-Joong Kim}.} \bibinfo{year}{2022}\natexlab{}.
\newblock \showarticletitle{Toward {Cooperative} {Level} {Generation} in {Multiplayer} {Games}: {A} {User} {Study} in {Overcooked}!}. In \bibinfo{booktitle}{\emph{2022 {IEEE} {Conference} on {Games} ({CoG})}}. \bibinfo{publisher}{IEEE}, \bibinfo{address}{New Jersey, USA}, \bibinfo{pages}{276--283}.
\newblock
\urldef\tempurl%
\url{https://doi.org/10.1109/CoG51982.2022.9893581}
\showDOI{\tempurl}
\newblock
\shownote{ISSN: 2325-4289}.


\bibitem[Bai et~al\mbox{.}(2021)]%
        {Bai_Hew_Gonda_2021}
\bibfield{author}{\bibinfo{person}{Shurui Bai}, \bibinfo{person}{Khe~Foon Hew}, {and} \bibinfo{person}{Donn~Emmanuel Gonda}.} \bibinfo{year}{2021}\natexlab{}.
\newblock \showarticletitle{Examining Effects of Different Leaderboards on Students’ Learning Performance, Intrinsic Motivation, and Perception in Gamified Online Learning Setting}. In \bibinfo{booktitle}{\emph{2021 IEEE International Conference on Educational Technology (ICET)}}. \bibinfo{publisher}{IEEE}, \bibinfo{address}{New Jersey, USA}, \bibinfo{pages}{36–41}.
\newblock
\urldef\tempurl%
\url{https://doi.org/10.1109/ICET52293.2021.9563130}
\showDOI{\tempurl}


\bibitem[Baumann et~al\mbox{.}(2024)]%
        {baumann_exploratory_2024}
\bibfield{author}{\bibinfo{person}{André C.~K. Baumann}, \bibinfo{person}{Ståle Pallesen}, \bibinfo{person}{Rune~A. Mentzoni}, \bibinfo{person}{Eirin Kolberg}, \bibinfo{person}{Vegard Waagbø}, \bibinfo{person}{Anders Sørensen}, {and} \bibinfo{person}{Joakim~H. Kristensen}.} \bibinfo{year}{2024}\natexlab{}.
\newblock \showarticletitle{An exploratory qualitative interview study on grassroots esports in sports clubs}.
\newblock \bibinfo{journal}{\emph{Frontiers in Sports and Active Living}}  \bibinfo{volume}{6} (\bibinfo{date}{Aug.} \bibinfo{year}{2024}), \bibinfo{numpages}{10}~pages.
\newblock
\showISSN{2624-9367}
\urldef\tempurl%
\url{https://doi.org/10.3389/fspor.2024.1405441}
\showDOI{\tempurl}


\bibitem[Bongaards et~al\mbox{.}(2024)]%
        {bongaards_personalized_2024}
\bibfield{author}{\bibinfo{person}{Thom Bongaards}, \bibinfo{person}{Maurits Adriaanse}, {and} \bibinfo{person}{Julian Frommel}.} \bibinfo{year}{2024}\natexlab{}.
\newblock \showarticletitle{Personalized {Matchmaking} {Restrictions} for {Reduced} {Exposure} to {Toxicity}: {Preliminary} {Insights} from an {Interview} {Study}}. In \bibinfo{booktitle}{\emph{Companion {Proceedings} of the 2024 {Annual} {Symposium} on {Computer}-{Human} {Interaction} in {Play}}} \emph{(\bibinfo{series}{{CHI} {PLAY} {Companion} '24})}. \bibinfo{publisher}{Association for Computing Machinery}, \bibinfo{address}{New York, NY, USA}, \bibinfo{pages}{31--36}.
\newblock
\showISBNx{9798400706929}
\urldef\tempurl%
\url{https://doi.org/10.1145/3665463.3678803}
\showDOI{\tempurl}


\bibitem[Bonilla~Gorrindo et~al\mbox{.}(2022a)]%
        {bonilla_gorrindo_psychological_2022}
\bibfield{author}{\bibinfo{person}{Ivan Bonilla~Gorrindo}, \bibinfo{person}{Andrés Chamarro}, {and} \bibinfo{person}{Carles Ventura}.} \bibinfo{year}{2022}\natexlab{a}.
\newblock \showarticletitle{Psychological skills in esports: {Qualitative} study of individual and team players}.
\newblock \bibinfo{journal}{\emph{Aloma: Revista de Psicologia, Ciències de l'Educació i de l'Esport}} \bibinfo{volume}{40}, \bibinfo{number}{1} (\bibinfo{date}{May} \bibinfo{year}{2022}), \bibinfo{pages}{35--41}.
\newblock
\showISSN{2339-9694, 1138-3194}
\urldef\tempurl%
\url{https://doi.org/10.51698/aloma.2022.40.1.36-41}
\showDOI{\tempurl}


\bibitem[Bonilla~Gorrindo et~al\mbox{.}(2022b)]%
        {Bonilla_Gorrindo_Chamarro_Ventura_2022}
\bibfield{author}{\bibinfo{person}{Ivan Bonilla~Gorrindo}, \bibinfo{person}{Andrés Chamarro}, {and} \bibinfo{person}{Carles Ventura}.} \bibinfo{year}{2022}\natexlab{b}.
\newblock \showarticletitle{Psychological skills in esports: Qualitative study of individual and team players}.
\newblock \bibinfo{journal}{\emph{Aloma: Revista de Psicologia, Ciències de l’Educació i de l’Esport}} \bibinfo{volume}{40}, \bibinfo{number}{1} (\bibinfo{date}{May} \bibinfo{year}{2022}), \bibinfo{pages}{35–41}.
\newblock
\showISSN{2339-9694, 1138-3194}
\urldef\tempurl%
\url{https://doi.org/10.51698/aloma.2022.40.1.36-41}
\showDOI{\tempurl}


\bibitem[Braun et~al\mbox{.}(2019)]%
        {Braun_Clarke_2019}
\bibfield{author}{\bibinfo{person}{Virginia Braun}, \bibinfo{person}{}, {and} \bibinfo{person}{Victoria Clarke}.} \bibinfo{year}{2019}\natexlab{}.
\newblock \showarticletitle{Reflecting on reflexive thematic analysis}.
\newblock \bibinfo{journal}{\emph{Qualitative Research in Sport, Exercise and Health}} \bibinfo{volume}{11}, \bibinfo{number}{4} (\bibinfo{date}{Aug.} \bibinfo{year}{2019}), \bibinfo{pages}{589–597}.
\newblock
\showISSN{2159-676X}
\urldef\tempurl%
\url{https://doi.org/10.1080/2159676X.2019.1628806}
\showDOI{\tempurl}


\bibitem[Braun and Clarke(2021)]%
        {Braun_Clarke_2021}
\bibfield{author}{\bibinfo{person}{Virginia Braun} {and} \bibinfo{person}{Victoria Clarke}.} \bibinfo{year}{2021}\natexlab{}.
\newblock \bibinfo{booktitle}{\emph{Thematic Analysis: A Practical Guide}}.
\newblock \bibinfo{publisher}{SAGE Publications}, \bibinfo{address}{New York, United States}.
\newblock
\showISBNx{978-1-5264-1730-5}


\bibitem[Breuer et~al\mbox{.}(2015)]%
        {breuer_sore_2015}
\bibfield{author}{\bibinfo{person}{Johannes Breuer}, \bibinfo{person}{Michael Scharkow}, {and} \bibinfo{person}{Thorsten Quandt}.} \bibinfo{year}{2015}\natexlab{}.
\newblock \showarticletitle{Sore losers? {A} reexamination of the frustration–aggression hypothesis for colocated video game play}.
\newblock \bibinfo{journal}{\emph{Psychology of Popular Media Culture}} \bibinfo{volume}{4}, \bibinfo{number}{2} (\bibinfo{year}{2015}), \bibinfo{pages}{126--137}.
\newblock
\showISSN{2160-4142}
\urldef\tempurl%
\url{https://doi.org/10.1037/ppm0000020}
\showDOI{\tempurl}


\bibitem[Bustamante and Spain(2008)]%
        {bustamante_measurement_2008}
\bibfield{author}{\bibinfo{person}{Ernesto~A. Bustamante} {and} \bibinfo{person}{Randall~D. Spain}.} \bibinfo{year}{2008}\natexlab{}.
\newblock \showarticletitle{Measurement {Invariance} of the {Nasa} {TLX}}.
\newblock \bibinfo{journal}{\emph{Proceedings of the Human Factors and Ergonomics Society Annual Meeting}} \bibinfo{volume}{52}, \bibinfo{number}{19} (\bibinfo{date}{Sept.} \bibinfo{year}{2008}), \bibinfo{pages}{1522--1526}.
\newblock
\showISSN{1071-1813, 2169-5067}
\urldef\tempurl%
\url{https://doi.org/10.1177/154193120805201946}
\showDOI{\tempurl}


\bibitem[Bányai et~al\mbox{.}(2019)]%
        {banyai_psychology_2019}
\bibfield{author}{\bibinfo{person}{Fanni Bányai}, \bibinfo{person}{Mark~D. Griffiths}, \bibinfo{person}{Orsolya Király}, {and} \bibinfo{person}{Zsolt Demetrovics}.} \bibinfo{year}{2019}\natexlab{}.
\newblock \showarticletitle{The {Psychology} of {Esports}: {A} {Systematic} {Literature} {Review}}.
\newblock \bibinfo{journal}{\emph{Journal of Gambling Studies}} \bibinfo{volume}{35}, \bibinfo{number}{2} (\bibinfo{date}{June} \bibinfo{year}{2019}), \bibinfo{pages}{351--365}.
\newblock
\showISSN{1573-3602}
\urldef\tempurl%
\url{https://doi.org/10.1007/s10899-018-9763-1}
\showDOI{\tempurl}


\bibitem[Cirett~Galán and Beal(2012)]%
        {cirett_galan_eeg_2012}
\bibfield{author}{\bibinfo{person}{Federico Cirett~Galán} {and} \bibinfo{person}{Carole~R. Beal}.} \bibinfo{year}{2012}\natexlab{}.
\newblock \showarticletitle{{EEG} {Estimates} of {Engagement} and {Cognitive} {Workload} {Predict} {Math} {Problem} {Solving} {Outcomes}}. In \bibinfo{booktitle}{\emph{User {Modeling}, {Adaptation}, and {Personalization}}}, \bibfield{editor}{\bibinfo{person}{Judith Masthoff}, \bibinfo{person}{Bamshad Mobasher}, \bibinfo{person}{Michel~C. Desmarais}, {and} \bibinfo{person}{Roger Nkambou}} (Eds.). \bibinfo{publisher}{Springer}, \bibinfo{address}{Berlin, Heidelberg}, \bibinfo{pages}{51--62}.
\newblock
\showISBNx{9783642314544}
\urldef\tempurl%
\url{https://doi.org/10.1007/978-3-642-31454-4_5}
\showDOI{\tempurl}


\bibitem[Drisko and Maschi(2016)]%
        {drisko2016content}
\bibfield{author}{\bibinfo{person}{James~W Drisko} {and} \bibinfo{person}{Tina Maschi}.} \bibinfo{year}{2016}\natexlab{}.
\newblock \bibinfo{booktitle}{\emph{Content analysis}}.
\newblock \bibinfo{publisher}{Oxford university press}, \bibinfo{address}{Oxford, United Kingdom}.
\newblock


\bibitem[Frommel and Mandryk(2024)]%
        {frommel_toxicity_2024}
\bibfield{author}{\bibinfo{person}{Julian Frommel} {and} \bibinfo{person}{Regan~L Mandryk}.} \bibinfo{year}{2024}\natexlab{}.
\newblock \showarticletitle{Toxicity in {Online} {Games}: {The} {Prevalence} and {Efficacy} of {Coping} {Strategies}}. In \bibinfo{booktitle}{\emph{Proceedings of the {CHI} {Conference} on {Human} {Factors} in {Computing} {Systems}}} \emph{(\bibinfo{series}{{CHI} '24})}. \bibinfo{publisher}{Association for Computing Machinery}, \bibinfo{address}{New York, NY, USA}, \bibinfo{pages}{1--12}.
\newblock
\showISBNx{9798400703300}
\urldef\tempurl%
\url{https://doi.org/10.1145/3613904.3642523}
\showDOI{\tempurl}


\bibitem[Gilleade and Dix(2004)]%
        {gilleade_using_2004}
\bibfield{author}{\bibinfo{person}{Kiel~M Gilleade} {and} \bibinfo{person}{Alan Dix}.} \bibinfo{year}{2004}\natexlab{}.
\newblock \showarticletitle{Using frustration in the design of adaptive videogames}. In \bibinfo{booktitle}{\emph{Proceedings of the 2004 {ACM} {SIGCHI} {International} {Conference} on {Advances} in computer entertainment technology}} \emph{(\bibinfo{series}{{ACE} '04})}. \bibinfo{publisher}{Association for Computing Machinery}, \bibinfo{address}{New York, NY, USA}, \bibinfo{pages}{228--232}.
\newblock
\showISBNx{9781581138825}
\urldef\tempurl%
\url{https://doi.org/10.1145/1067343.1067372}
\showDOI{\tempurl}


\bibitem[Gisbert-Pérez et~al\mbox{.}(2024)]%
        {gisbert-perez_key_2024}
\bibfield{author}{\bibinfo{person}{Júlia Gisbert-Pérez}, \bibinfo{person}{Alejo García-Naveira}, \bibinfo{person}{Manuel Martí-Vilar}, {and} \bibinfo{person}{Jorge Acebes-Sánchez}.} \bibinfo{year}{2024}\natexlab{}.
\newblock \showarticletitle{Key structure and processes in esports teams: a systematic review}.
\newblock \bibinfo{journal}{\emph{Current Psychology}} \bibinfo{volume}{43}, \bibinfo{number}{23} (\bibinfo{date}{June} \bibinfo{year}{2024}), \bibinfo{pages}{20355--20374}.
\newblock
\showISSN{1936-4733}
\urldef\tempurl%
\url{https://doi.org/10.1007/s12144-024-05858-0}
\showDOI{\tempurl}


\bibitem[Greitemeyer and Cox(2013)]%
        {greitemeyer_theres_2013}
\bibfield{author}{\bibinfo{person}{Tobias Greitemeyer} {and} \bibinfo{person}{Christopher Cox}.} \bibinfo{year}{2013}\natexlab{}.
\newblock \showarticletitle{There's no “{I}” in team: {Effects} of cooperative video games on cooperative behavior}.
\newblock \bibinfo{journal}{\emph{European Journal of Social Psychology}} \bibinfo{volume}{43}, \bibinfo{number}{3} (\bibinfo{date}{April} \bibinfo{year}{2013}), \bibinfo{pages}{224--228}.
\newblock
\showISSN{0046-2772, 1099-0992}
\urldef\tempurl%
\url{https://doi.org/10.1002/ejsp.1940}
\showDOI{\tempurl}


\bibitem[Hanghøj et~al\mbox{.}(2018)]%
        {hanghoj_can_2018}
\bibfield{author}{\bibinfo{person}{Thorkild Hanghøj}, \bibinfo{person}{Andreas Lieberoth}, {and} \bibinfo{person}{Morten Misfeldt}.} \bibinfo{year}{2018}\natexlab{}.
\newblock \showarticletitle{Can cooperative video games encourage social and motivational inclusion of at‐risk students?}
\newblock \bibinfo{journal}{\emph{British Journal of Educational Technology}} \bibinfo{volume}{49}, \bibinfo{number}{4} (\bibinfo{date}{July} \bibinfo{year}{2018}), \bibinfo{pages}{775--799}.
\newblock
\showISSN{0007-1013, 1467-8535}
\urldef\tempurl%
\url{https://doi.org/10.1111/bjet.12642}
\showDOI{\tempurl}


\bibitem[Hanghøj and Nielsen(2019)]%
        {hanghoj_esports_2019}
\bibfield{author}{\bibinfo{person}{Thorkild Hanghøj} {and} \bibinfo{person}{Rune Nielsen}.} \bibinfo{year}{2019}\natexlab{}.
\newblock \showarticletitle{{eSports} {Skills} are {People} {Skills}}. In \bibinfo{booktitle}{\emph{"{Proceedings} of the 12th {European} {Conference} on {Game} {Based} {Learning} "}}. \bibinfo{publisher}{ACPI}, \bibinfo{address}{Sophia Antipolis, France}, \bibinfo{pages}{63}.
\newblock
\showISBNx{9781912764389}
\urldef\tempurl%
\url{https://doi.org/10.34190/GBL.19.041}
\showDOI{\tempurl}


\bibitem[Hart and Field(2006)]%
        {Hart_Field}
\bibfield{author}{\bibinfo{person}{Sandra~G Hart} {and} \bibinfo{person}{Moffett Field}.} \bibinfo{year}{2006}\natexlab{}.
\newblock \showarticletitle{NASA-TASK LOAD INDEX (NASA-TLX); 20 YEARS LATER}. In \bibinfo{booktitle}{\emph{Proceedings of the Human Factors and Ergonomics Society 50th Annual Meeting}}. \bibinfo{publisher}{SAGE}, \bibinfo{address}{Santa Monica}, \bibinfo{pages}{904--908}.
\newblock
\urldef\tempurl%
\url{https://humansystems.arc.nasa.gov/groups/tlx/}
\showURL{%
\tempurl}


\bibitem[Kukshinov et~al\mbox{.}(2025)]%
        {Kukshinov_Tu_Szita_SenthilNathan_Nacke_2025}
\bibfield{author}{\bibinfo{person}{Eugene Kukshinov}, \bibinfo{person}{Joseph Tu}, \bibinfo{person}{Kata Szita}, \bibinfo{person}{Kaushall Senthil~Nathan}, {and} \bibinfo{person}{Lennart~E Nacke}.} \bibinfo{year}{2025}\natexlab{}.
\newblock \showarticletitle{Widespread yet Unreliable: A Systematic Analysis of the Use of Presence Questionnaires}.
\newblock \bibinfo{journal}{\emph{Interacting with Computers}} \bibinfo{volume}{1}, \bibinfo{number}{1} (\bibinfo{date}{Feb.} \bibinfo{year}{2025}), \bibinfo{numpages}{16}~pages.
\newblock
\showISSN{1873-7951}
\urldef\tempurl%
\url{https://doi.org/10.1093/iwc/iwae064}
\showDOI{\tempurl}


\bibitem[Lee et~al\mbox{.}(2024)]%
        {lee_characterizing_2024}
\bibfield{author}{\bibinfo{person}{Hanbyeol Lee}, \bibinfo{person}{Seyeon Lee}, \bibinfo{person}{Rohan Nallapati}, \bibinfo{person}{Youngjung Uh}, {and} \bibinfo{person}{Byungjoo Lee}.} \bibinfo{year}{2024}\natexlab{}.
\newblock \showarticletitle{Characterizing and {Quantifying} {Expert} {Input} {Behavior} in {League} of {Legends}}. In \bibinfo{booktitle}{\emph{Proceedings of the {CHI} {Conference} on {Human} {Factors} in {Computing} {Systems}}}. \bibinfo{publisher}{ACM}, \bibinfo{address}{Honolulu HI USA}, \bibinfo{pages}{1--21}.
\newblock
\showISBNx{9798400703300}
\urldef\tempurl%
\url{https://doi.org/10.1145/3613904.3642588}
\showDOI{\tempurl}


\bibitem[Maharani et~al\mbox{.}(2024)]%
        {maharani_understanding_2024}
\bibfield{author}{\bibinfo{person}{Anissa Maharani}, \bibinfo{person}{Virienia Puspita}, \bibinfo{person}{Raden~Arny Aurora}, {and} \bibinfo{person}{Nico Wiranito}.} \bibinfo{year}{2024}\natexlab{}.
\newblock \showarticletitle{Understanding {Toxicity} in {Online} {Gaming}: {A} {Focus} on {Communication}-{Based} {Behaviours} towards {Female} {Players} in {Valorant}}.
\newblock \bibinfo{journal}{\emph{Jurnal Syntax Admiration}} \bibinfo{volume}{5}, \bibinfo{number}{5} (\bibinfo{date}{May} \bibinfo{year}{2024}), \bibinfo{pages}{1559--1567}.
\newblock
\showISSN{2722-5356}
\urldef\tempurl%
\url{https://doi.org/10.46799/jsa.v5i5.1137}
\showDOI{\tempurl}


\bibitem[Martončik(2015)]%
        {martoncik_e-sports_2015}
\bibfield{author}{\bibinfo{person}{Marcel Martončik}.} \bibinfo{year}{2015}\natexlab{}.
\newblock \showarticletitle{e-{Sports}: {Playing} just for fun or playing to satisfy life goals?}
\newblock \bibinfo{journal}{\emph{Computers in Human Behavior}}  \bibinfo{volume}{48} (\bibinfo{date}{July} \bibinfo{year}{2015}), \bibinfo{pages}{208--211}.
\newblock
\showISSN{0747-5632}
\urldef\tempurl%
\url{https://doi.org/10.1016/j.chb.2015.01.056}
\showDOI{\tempurl}


\bibitem[McLean et~al\mbox{.}(2020)]%
        {mclean_toxic_2020}
\bibfield{author}{\bibinfo{person}{Dave McLean}, \bibinfo{person}{Frank Waddell}, {and} \bibinfo{person}{James Ivory}.} \bibinfo{year}{2020}\natexlab{}.
\newblock \showarticletitle{Toxic {Teammates} or {Obscene} {Opponents}? {Influences} of {Cooperation} and {Competition} on {Hostility} between {Teammates} and {Opponents} in an {Online} {Game}}.
\newblock \bibinfo{journal}{\emph{Journal For Virtual Worlds Research}} \bibinfo{volume}{13}, \bibinfo{number}{1} (\bibinfo{date}{March} \bibinfo{year}{2020}), \bibinfo{numpages}{17}~pages.
\newblock
\showISSN{1941-8477}
\urldef\tempurl%
\url{https://doi.org/10.4101/jvwr.v13i1.7334}
\showDOI{\tempurl}


\bibitem[Musick et~al\mbox{.}(2021)]%
        {musick_leveling_2021}
\bibfield{author}{\bibinfo{person}{Geoff Musick}, \bibinfo{person}{Rui Zhang}, \bibinfo{person}{Nathan~J. McNeese}, \bibinfo{person}{Guo Freeman}, {and} \bibinfo{person}{Anurata~Prabha Hridi}.} \bibinfo{year}{2021}\natexlab{}.
\newblock \showarticletitle{Leveling {Up} {Teamwork} in {Esports}: {Understanding} {Team} {Cognition} in a {Dynamic} {Virtual} {Environment}}.
\newblock \bibinfo{journal}{\emph{Proceedings of the ACM on Human-Computer Interaction}} \bibinfo{volume}{5}, \bibinfo{number}{CSCW1} (\bibinfo{date}{April} \bibinfo{year}{2021}), \bibinfo{pages}{1--30}.
\newblock
\showISSN{2573-0142}
\urldef\tempurl%
\url{https://doi.org/10.1145/3449123}
\showDOI{\tempurl}


\bibitem[Nagorsky and Wiemeyer(2020)]%
        {nagorsky_structure_2020}
\bibfield{author}{\bibinfo{person}{Eugen Nagorsky} {and} \bibinfo{person}{Josef Wiemeyer}.} \bibinfo{year}{2020}\natexlab{}.
\newblock \showarticletitle{The structure of performance and training in esports}.
\newblock \bibinfo{journal}{\emph{PLOS ONE}} \bibinfo{volume}{15}, \bibinfo{number}{8} (\bibinfo{date}{Aug.} \bibinfo{year}{2020}), \bibinfo{pages}{e0237584}.
\newblock
\showISSN{1932-6203}
\urldef\tempurl%
\url{https://doi.org/10.1371/journal.pone.0237584}
\showDOI{\tempurl}


\bibitem[Neri et~al\mbox{.}(2021)]%
        {neri_personalized_2021}
\bibfield{author}{\bibinfo{person}{Francesco Neri}, \bibinfo{person}{Carmelo~Luca Smeralda}, \bibinfo{person}{Davide Momi}, \bibinfo{person}{Giulia Sprugnoli}, \bibinfo{person}{Arianna Menardi}, \bibinfo{person}{Salvatore Ferrone}, \bibinfo{person}{Simone Rossi}, \bibinfo{person}{Alessandro Rossi}, \bibinfo{person}{Giorgio Di~Lorenzo}, {and} \bibinfo{person}{Emiliano Santarnecchi}.} \bibinfo{year}{2021}\natexlab{}.
\newblock \showarticletitle{Personalized {Adaptive} {Training} {Improves} {Performance} at a {Professional} {First}-{Person} {Shooter} {Action} {Videogame}}.
\newblock \bibinfo{journal}{\emph{Frontiers in Psychology}}  \bibinfo{volume}{12} (\bibinfo{date}{June} \bibinfo{year}{2021}), \bibinfo{pages}{598410}.
\newblock
\showISSN{1664-1078}
\urldef\tempurl%
\url{https://doi.org/10.3389/fpsyg.2021.598410}
\showDOI{\tempurl}


\bibitem[Nexø(2024)]%
        {nexo_toxic_2024}
\bibfield{author}{\bibinfo{person}{Louise~Anker Nexø}.} \bibinfo{year}{2024}\natexlab{}.
\newblock \showarticletitle{Toxic {Behaviours} in {Esport}: {A} {Review} of {Data}-{Collection} {Methods} {Applied} in {Studying} {Toxic} {In}-{Gaming} {Behaviours}}.
\newblock \bibinfo{journal}{\emph{International Journal of Esports}} \bibinfo{volume}{3}, \bibinfo{number}{3} (\bibinfo{date}{Jan.} \bibinfo{year}{2024}), \bibinfo{numpages}{28}~pages.
\newblock
\showISSN{2634-1069}
\urldef\tempurl%
\url{https://www.ijesports.org/article/127/html}
\showURL{%
\tempurl}


\bibitem[Perrig et~al\mbox{.}(2024)]%
        {Perrig_Aeschbach_Scharowski_vonFelten_Opwis_Brühlmann_2024}
\bibfield{author}{\bibinfo{person}{Sebastian A.~C. Perrig}, \bibinfo{person}{Lena~Fanya Aeschbach}, \bibinfo{person}{Nicolas Scharowski}, \bibinfo{person}{Nick von Felten}, \bibinfo{person}{Klaus Opwis}, {and} \bibinfo{person}{Florian Brühlmann}.} \bibinfo{year}{2024}\natexlab{}.
\newblock \showarticletitle{Measurement practices in user experience (UX) research: a systematic quantitative literature review}.
\newblock \bibinfo{journal}{\emph{Frontiers in Computer Science}}  \bibinfo{volume}{6} (\bibinfo{date}{March} \bibinfo{year}{2024}), \bibinfo{numpages}{17}~pages.
\newblock
\showISSN{2624-9898}
\urldef\tempurl%
\url{https://doi.org/10.3389/fcomp.2024.1368860}
\showDOI{\tempurl}


\bibitem[Pizzo et~al\mbox{.}(2022)]%
        {pizzo_esports_2022}
\bibfield{author}{\bibinfo{person}{Anthony~D. Pizzo}, \bibinfo{person}{Yiran Su}, \bibinfo{person}{Tobias Scholz}, \bibinfo{person}{Bradley~J. Baker}, \bibinfo{person}{Juho Hamari}, {and} \bibinfo{person}{Leah Ndanga}.} \bibinfo{year}{2022}\natexlab{}.
\newblock \showarticletitle{Esports {Scholarship} {Review}: {Synthesis}, {Contributions}, and {Future} {Research}}.
\newblock \bibinfo{journal}{\emph{Journal of Sport Management}} \bibinfo{volume}{36}, \bibinfo{number}{3} (\bibinfo{date}{May} \bibinfo{year}{2022}), \bibinfo{pages}{228--239}.
\newblock
\showISSN{0888-4773, 1543-270X}
\urldef\tempurl%
\url{https://doi.org/10.1123/jsm.2021-0228}
\showDOI{\tempurl}


\bibitem[Reid et~al\mbox{.}(2022)]%
        {reid_feeling_2022}
\bibfield{author}{\bibinfo{person}{Elizabeth Reid}, \bibinfo{person}{Regan~L. Mandryk}, \bibinfo{person}{Nicole~A. Beres}, \bibinfo{person}{Madison Klarkowski}, {and} \bibinfo{person}{Julian Frommel}.} \bibinfo{year}{2022}\natexlab{}.
\newblock \showarticletitle{Feeling {Good} and {In} {Control}: {In}-game {Tools} to {Support} {Targets} of {Toxicity}}.
\newblock \bibinfo{journal}{\emph{Proc. ACM Hum.-Comput. Interact.}} \bibinfo{volume}{6}, \bibinfo{number}{CHI PLAY} (\bibinfo{date}{Oct.} \bibinfo{year}{2022}), \bibinfo{pages}{235:1--235:27}.
\newblock
\urldef\tempurl%
\url{https://doi.org/10.1145/3549498}
\showDOI{\tempurl}


\bibitem[Sabtan et~al\mbox{.}(2022)]%
        {sabtan_current_2022}
\bibfield{author}{\bibinfo{person}{Bader Sabtan}, \bibinfo{person}{Shi Cao}, {and} \bibinfo{person}{Naomi Paul}.} \bibinfo{year}{2022}\natexlab{}.
\newblock \showarticletitle{Current practice and challenges in coaching {Esports} players: {An} interview study with league of legends professional team coaches}.
\newblock \bibinfo{journal}{\emph{Entertainment Computing}}  \bibinfo{volume}{42} (\bibinfo{date}{May} \bibinfo{year}{2022}), \bibinfo{pages}{100481}.
\newblock
\showISSN{1875-9521}
\urldef\tempurl%
\url{https://doi.org/10.1016/j.entcom.2022.100481}
\showDOI{\tempurl}


\bibitem[Sachan et~al\mbox{.}(2025)]%
        {sachan_social_2025}
\bibfield{author}{\bibinfo{person}{Tushya Sachan}, \bibinfo{person}{Dinesh Chhabra}, {and} \bibinfo{person}{Betina Abraham}.} \bibinfo{year}{2025}\natexlab{}.
\newblock \showarticletitle{Social {Capital} in {Online} {Gaming} {Communities}: {A} {Systematic} {Review} {Examining} the {Role} of {Virtual} {Identities}}.
\newblock \bibinfo{journal}{\emph{Cyberpsychology, Behavior, and Social Networking}} \bibinfo{volume}{28}, \bibinfo{number}{3} (\bibinfo{date}{March} \bibinfo{year}{2025}), \bibinfo{pages}{147--161}.
\newblock
\showISSN{2152-2715, 2152-2723}
\urldef\tempurl%
\url{https://doi.org/10.1089/cyber.2024.0375}
\showDOI{\tempurl}


\bibitem[Seif El-Nasr et~al\mbox{.}(2010)]%
        {seif_el-nasr_understanding_2010}
\bibfield{author}{\bibinfo{person}{Magy Seif El-Nasr}, \bibinfo{person}{Bardia Aghabeigi}, \bibinfo{person}{David Milam}, \bibinfo{person}{Mona Erfani}, \bibinfo{person}{Beth Lameman}, \bibinfo{person}{Hamid Maygoli}, {and} \bibinfo{person}{Sang Mah}.} \bibinfo{year}{2010}\natexlab{}.
\newblock \showarticletitle{Understanding and evaluating cooperative games}. In \bibinfo{booktitle}{\emph{Proceedings of the {SIGCHI} {Conference} on {Human} {Factors} in {Computing} {Systems}}}. \bibinfo{publisher}{ACM}, \bibinfo{address}{Atlanta Georgia USA}, \bibinfo{pages}{253--262}.
\newblock
\showISBNx{9781605589299}
\urldef\tempurl%
\url{https://doi.org/10.1145/1753326.1753363}
\showDOI{\tempurl}


\bibitem[Shen et~al\mbox{.}(2020)]%
        {shen_viral_2020}
\bibfield{author}{\bibinfo{person}{Cuihua Shen}, \bibinfo{person}{Qiusi Sun}, \bibinfo{person}{Taeyoung Kim}, \bibinfo{person}{Grace Wolff}, \bibinfo{person}{Rabindra Ratan}, {and} \bibinfo{person}{Dmitri Williams}.} \bibinfo{year}{2020}\natexlab{}.
\newblock \showarticletitle{Viral vitriol: {Predictors} and contagion of online toxicity in {World} of {Tanks}}.
\newblock \bibinfo{journal}{\emph{Computers in Human Behavior}}  \bibinfo{volume}{108} (\bibinfo{date}{July} \bibinfo{year}{2020}), \bibinfo{pages}{106343}.
\newblock
\showISSN{0747-5632}
\urldef\tempurl%
\url{https://doi.org/10.1016/j.chb.2020.106343}
\showDOI{\tempurl}


\bibitem[Studios(2021)]%
        {HazelightStudios_2021}
\bibfield{author}{\bibinfo{person}{Hazelight Studios}.} \bibinfo{year}{2021}\natexlab{}.
\newblock \bibinfo{title}{It Takes Two}.
\newblock
\newblock
\urldef\tempurl%
\url{https://www.ea.com/games/it-takes-two}
\showURL{%
\tempurl}


\bibitem[Team17(2018)]%
        {noauthor_overcooked_nodate}
\bibfield{author}{\bibinfo{person}{Team17}.} \bibinfo{year}{2018}\natexlab{}.
\newblock \bibinfo{title}{Overcooked {\textbar} {Cooking} {Video} {Game} {\textbar} {Team17}}.
\newblock
\newblock
\urldef\tempurl%
\url{https://www.team17.com/games/overcooked}
\showURL{%
\tempurl}


\bibitem[Trepanowski et~al\mbox{.}(2024)]%
        {trepanowski_sexism_2024}
\bibfield{author}{\bibinfo{person}{Radosław Trepanowski}, \bibinfo{person}{Samuli Laato}, \bibinfo{person}{Dariusz Drążkowski}, \bibinfo{person}{Juho Hamari}, {and} \bibinfo{person}{Zuzanna Kopeć}.} \bibinfo{year}{2024}\natexlab{}.
\newblock \showarticletitle{Sexism in esports: {How} male and female players evaluate each others’ performance and agency}.
\newblock \bibinfo{journal}{\emph{Computers in Human Behavior}}  \bibinfo{volume}{161} (\bibinfo{date}{Dec.} \bibinfo{year}{2024}), \bibinfo{pages}{108415}.
\newblock
\showISSN{0747-5632}
\urldef\tempurl%
\url{https://doi.org/10.1016/j.chb.2024.108415}
\showDOI{\tempurl}


\bibitem[Trotter et~al\mbox{.}(2021)]%
        {trotter_social_2021}
\bibfield{author}{\bibinfo{person}{Michael~G. Trotter}, \bibinfo{person}{Tristan~J. Coulter}, \bibinfo{person}{Paul~A. Davis}, \bibinfo{person}{Dylan~R. Poulus}, {and} \bibinfo{person}{Remco Polman}.} \bibinfo{year}{2021}\natexlab{}.
\newblock \showarticletitle{Social {Support}, {Self}-{Regulation}, and {Psychological} {Skill} {Use} in {E}-{Athletes}}.
\newblock \bibinfo{journal}{\emph{Frontiers in Psychology}}  \bibinfo{volume}{12} (\bibinfo{date}{Nov.} \bibinfo{year}{2021}), \bibinfo{numpages}{10}~pages.
\newblock
\showISSN{1664-1078}
\urldef\tempurl%
\url{https://doi.org/10.3389/fpsyg.2021.722030}
\showDOI{\tempurl}


\bibitem[Wang et~al\mbox{.}(2025)]%
        {Wang_2025}
\bibfield{author}{\bibinfo{person}{Derrick~M. Wang}, \bibinfo{person}{Sebastian Cmentowski}, \bibinfo{person}{Reza Hadi~Mogavi}, \bibinfo{person}{Kaushall Senthil~Nathan}, \bibinfo{person}{Eugene Kukshinov}, \bibinfo{person}{Joseph Tu}, {and} \bibinfo{person}{Lennart~E. Nacke}.} \bibinfo{year}{2025}\natexlab{}.
\newblock \showarticletitle{From Solo to Social: Exploring the Dynamics of Player Cooperation in a Co-located Cooperative Exergame}. In \bibinfo{booktitle}{\emph{Proceedings of the 2025 CHI Conference on Human Factors in Computing Systems}} \emph{(\bibinfo{series}{CHI ’25})}. \bibinfo{publisher}{Association for Computing Machinery}, \bibinfo{address}{New York, NY, USA}, \bibinfo{pages}{1–16}.
\newblock
\showISBNx{979-8-4007-1394-1}
\urldef\tempurl%
\url{https://doi.org/10.1145/3706598.3713937}
\showDOI{\tempurl}


\bibitem[Webster et~al\mbox{.}(2014)]%
        {webster_brief_2014}
\bibfield{author}{\bibinfo{person}{Gregory~D. Webster}, \bibinfo{person}{C.~Nathan DeWall}, \bibinfo{person}{Richard~S. Pond}, \bibinfo{person}{Timothy Deckman}, \bibinfo{person}{Peter~K. Jonason}, \bibinfo{person}{Bonnie~M. Le}, \bibinfo{person}{Austin~Lee Nichols}, \bibinfo{person}{Tatiana~Orozco Schember}, \bibinfo{person}{Laura~C. Crysel}, \bibinfo{person}{Benjamin~S. Crosier}, \bibinfo{person}{C.~Veronica Smith}, \bibinfo{person}{E.~Layne Paddock}, \bibinfo{person}{John~B. Nezlek}, \bibinfo{person}{Lee~A. Kirkpatrick}, \bibinfo{person}{Angela~D. Bryan}, {and} \bibinfo{person}{Renée~J. Bator}.} \bibinfo{year}{2014}\natexlab{}.
\newblock \showarticletitle{The brief aggression questionnaire: psychometric and behavioral evidence for an efficient measure of trait aggression}.
\newblock \bibinfo{journal}{\emph{Aggressive Behavior}} \bibinfo{volume}{40}, \bibinfo{number}{2} (\bibinfo{date}{March} \bibinfo{year}{2014}), \bibinfo{pages}{120--139}.
\newblock
\showISSN{0096-140X, 1098-2337}
\urldef\tempurl%
\url{https://doi.org/10.1002/ab.21507}
\showDOI{\tempurl}


\bibitem[Wijkstra(2024)]%
        {wijkstra_fighting_2024}
\bibfield{author}{\bibinfo{person}{Michel Wijkstra}.} \bibinfo{year}{2024}\natexlab{}.
\newblock \showarticletitle{Fighting {Toxicity} {Through} {Positive} and {Preventative} {Intervention}}. In \bibinfo{booktitle}{\emph{Companion {Proceedings} of the 2024 {Annual} {Symposium} on {Computer}-{Human} {Interaction} in {Play}}} \emph{(\bibinfo{series}{{CHI} {PLAY} {Companion} '24})}. \bibinfo{publisher}{Association for Computing Machinery}, \bibinfo{address}{New York, NY, USA}, \bibinfo{pages}{450--453}.
\newblock
\showISBNx{9798400706929}
\urldef\tempurl%
\url{https://doi.org/10.1145/3665463.3678857}
\showDOI{\tempurl}


\bibitem[Yee et~al\mbox{.}(2012)]%
        {10.1145/2207676.2208681}
\bibfield{author}{\bibinfo{person}{Nick Yee}, \bibinfo{person}{Nicolas Ducheneaut}, {and} \bibinfo{person}{Les Nelson}.} \bibinfo{year}{2012}\natexlab{}.
\newblock \showarticletitle{Online gaming motivations scale: development and validation}. In \bibinfo{booktitle}{\emph{Proceedings of the SIGCHI Conference on Human Factors in Computing Systems}} (Austin, Texas, USA) \emph{(\bibinfo{series}{CHI '12})}. \bibinfo{publisher}{Association for Computing Machinery}, \bibinfo{address}{New York, NY, USA}, \bibinfo{pages}{2803–2806}.
\newblock
\showISBNx{9781450310154}
\urldef\tempurl%
\url{https://doi.org/10.1145/2207676.2208681}
\showDOI{\tempurl}


\bibitem[Zhang et~al\mbox{.}(2024)]%
        {zhang_toxicity_2024}
\bibfield{author}{\bibinfo{person}{Zinan Zhang}, \bibinfo{person}{Sam Moradzadeh}, \bibinfo{person}{Andrew Woan}, {and} \bibinfo{person}{Yubo Kou}.} \bibinfo{year}{2024}\natexlab{}.
\newblock \showarticletitle{Toxicity by {Game} {Design}: {How} {Players} {Perceive} the {Influence} of {Game} {Design} on {Toxicity}}.
\newblock \bibinfo{journal}{\emph{Proc. ACM Hum.-Comput. Interact.}} \bibinfo{volume}{8}, \bibinfo{number}{CHI PLAY} (\bibinfo{date}{Oct.} \bibinfo{year}{2024}), \bibinfo{pages}{345:1--345:31}.
\newblock
\urldef\tempurl%
\url{https://doi.org/10.1145/3677110}
\showDOI{\tempurl}


\end{thebibliography}
